# Quantization of Charge Carriers in Conduction Channels of Si-Based Field-Effect Transistors for Multinary Computation

P. Xu and H. Luo*

*Department of Physics, University at Buffalo, The State University of New York, NY*

*14260, USA*

The latest field-effect transistors are entering the regime where quantum effects within the conduction channel can play a significant role because of the increasingly reduced dimensions. We investigate the effects of quantized states in conduction channels in transistors with dimensions close to those presently used. We use the standard configuration of Si-based metal-oxide-semiconductor field-effect transistors (MOSFETs), as a simplified model to provide an estimate of the effect of quantization with respect to the dimensions of the conduction channel. The study shows simulated results of drain currents for various combinations of dimensions, in which distinguishable current levels as a function of the applied gate bias can be obtained at room temperature. The same qualitative dependence on dimensions is expected to apply to the state-of-the-art transistor architectures with dimensions near this range, such as fin field-effect transistors (FinFETs) and gate-all-around field-effect transistors (GAAFETs). The results show that utilizing quantized states in the conduction channel for multinary computation has become a possibility with their present dimensions.

The advances in semiconductor industries led to a continuous decrease in device feature size. With improvements in design and manufacturing, it is close to the point where quantization of electrons/holes within the conduction channel can lead to new opportunities and problems. The focus so far has been primarily on increasing the area density of transistors, and overcoming complexities resulting from reduced dimensions, such as short channel effects. While multinary transistors have not been a main focus in the field of semiconductor physics and applications as a whole, there have been many attempts, to investigate both multinary computation and the corresponding devices. For example, ternary computers have been studied as early as 1950s.[1] Great success was achieved with floating gate memory devices with the wide spread 3D structures.[2] But the same concept proved to be difficult to implement for processors with a different gating method. It is in principle possible to break up the range of the operating gate voltage into multiple regions and use them as multinary states. But for a current curve without quantized steps, the tolerance for voltage variations makes it not practical. This will be further discussed later. More recently, there have been studies of ternary states, as opposed to the commonly used binary states, by using band-to-band tunneling with two transistors of different charge carriers integrated, which showed promise in this direction.[3] However, those approaches involve more than one transistor for each bit, a disadvantage for increasing area density of bits. Quantum computing is the ultimate use of quantized states, which currently involves extremely low temperatures. For mobile devices, however, field-effect transistors will likely continue to be the choice under ambient conditions. Currently, state-of-the-art field-effect transistors on the nanometer scale have evolved into more complex 3D structures, such as fin field-effect transistors (FinFETs) and gate-all-around field-effect transistors (GAAFETs), to overcome primarily the short channel effects with large leakage currents, among other problems, related to the reduced dimensions.[4] In both FinFETs and GAAFETs, their dimensions are approaching the regime where the separations of the quantized energy levels exceed the thermal energy. The main efforts in moving to smaller dimensions have been devoted to minimizing the effects of the source and the drain on the channel. In the context of this work, isolation of the channel from the source and the drain corresponds to an improvement of the confinement of charge carriers in the channel. Unfortunately, fully quantum mechanical approaches for the advanced transistor architectures remain challenging because of the complexity and variety of the structures, and often charge-dependent potential profiles. For simplicity and generality, we will focus on the common characteristics of quantization associated with nanometer dimensions, using the conventional MOSFET configuration as our idealized model, for which many analytical solutions and established methods are available. The qualitative behavior of quantization of the charge carriers on the nanometer scale is, however, expected to be similar among different designs, as seen in nanostructures of different shapes and materials. For the purpose of this study, the short channel effects on the nanometer scale MOSFETs, which have been greatly reduced in other state-of-the-art designs such as FinFETs, are not included.

*Corresponding Author. E-mail address: luo@buffalo.edu.

One of the fundamental challenges involved in the use of quantized states as the basis for conduction features, such as steps as a function of the gate bias, is the energy separation between those states in comparison with the thermal energy, *kT*, where *k* is the Boltzmann constant and *T* is the temperature. Earlier studies of quantum wires in the configuration of transistors have shown very desirable features, with step-like currents with increasing applied gate voltage.[5] However, such features are only visible at low temperatures because the width and the length are large and the energy separations of the quantized states are small. In this study, we use the conventional Si-based MOSFET structure and investigate the effects of the conduction channel, near practical dimensions for the most advanced transistors. This allows the use of the quantized states for creating step-like features in the current as a function of the applied gate bias, most importantly, at room temperature. Moreover, the multinary transistors discussed here do not require architectures with more terminals than the present three terminals.

The idealized MOSFET is shown in Fig. 1.

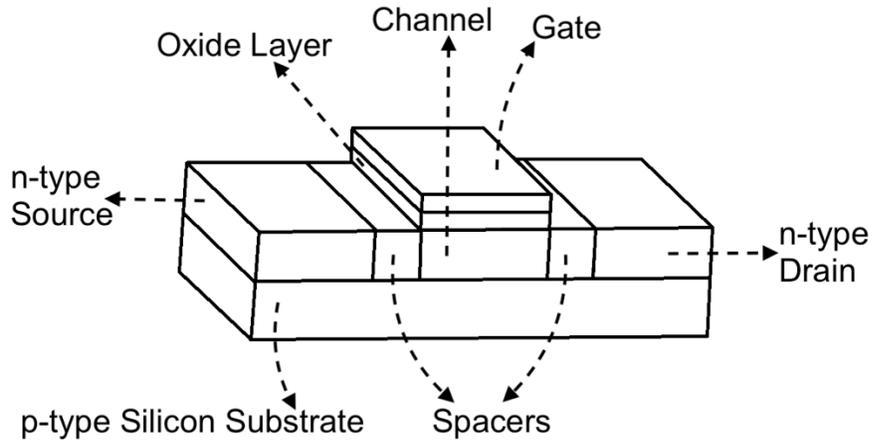

Fig. 1. Schematic diagram of a MOSFET structure as an example to demonstrate the effect of quantization of electron states on device functionality.

In this structure, the gate does not cover the whole conduction channel, leaving two ungated spacers between the gate and the source/drain regions, as used in some designs. We use the following parameters for the simulation. The substrate is p-type silicon with a dopant concentration of $N_A = 10^{18}\ cm^{-3}$. The oxide layer thickness is $t_{ox} = 1\ nm$, in the range of some of the currently used devices. For the same oxide layer thickness, higher dielectric constants will lead to stronger energy dependence on the electric fields within the inversion layer associated with the applied gate bias, and thus more changes of the quantized energy levels of the inversion electrons. A dielectric constant reported by InTel is used, with high-k dielectrics such as Hafnium Oxide, $H_f O_2$. The dielectric constant of $H_f O_2$ is between 22-25.[6] The value used in this study is 22.

A quantum mechanical treatment of the inversion charge $Q_{in}$ has been done before by solving the coupled Poisson and Schrodinger equations.[7,8] For simplicity, a commonly used approximation is to have a triangular well in the inversion layer.[9] The slope and the bottom of the triangular well require self-consistency in connection with the inversion charge. Because they are critical to the simulation, they are calculated self-consistently.

The quantized energy levels in the triangular well are given by:[10]

$$E_n = \left[\frac{3}{2}\pi\left(n - \frac{1}{4}\right)\right]^{\frac{2}{3}} \left[\frac{(eF\hbar)^2}{2m_z}\right]^{\frac{1}{3}}, \tag{1}$$

with $n = 1, 2, 3, \ldots$ The slope of the triangular well is $eF$, where $e$ is the absolute value of the electron charge, and $F$ the electric field which will be determined later. In earlier studies,[8, 11] a 2D density of states is assumed for the conduction channel. This allows for an analytical expression for the inversion charge $Q_{in}$. However, in the MOSFETs used here, the channel length and width can be less than 10 nm, a 2D density of states cannot be used. Instead, the density of states in the inversion layer is determined quantum mechanically, which is used for the self-consistent calculation.

For the effective mass $m^*$ in Si, the six-fold degenerate conduction band ellipsoids are considered, which are surfaces of constant energy in k-space. They are described by two effective masses, the longitudinal effective mass $m_l = 0.98\, m_e$, and the transverse effective mass $m_t = 0.19\, m_e$, along the major and the minor axes of the ellipsoids, respectively.[12] In our simulation, the gate electric field is along the [100] direction of the silicon substrate. We choose the channel length direction to be along the [110] direction, as is often the case in practice. The corresponding channel width is along the [1$\bar{1}$0] direction. The energies of the quantized states are calculated with the corresponding effective masses that can be derived from the ellipsoids.

The approach described below are simplified, but is expected to be indicative of the effect of quantization in the nanometer range. In the direction of the gate field, we use the triangular well approximation, as mentioned before. Along the direction of the width, we treat the channel as an infinite quantum well, similar to some treatments used for quantum wire transistors. For the channel length direction, quantization is not a necessary condition for step-like features as shown in the experimental results of quantum wire transistors.[5] Although the channel is a barrier in this direction at zero gate bias, it does not eliminate the presence of some degree of confinement effect, as has been observed for quasi-localized states above quantum barriers, related to the abrupt potential changes at the interfaces.[13] We assume that there is some level of quantization along the channel length direction, because of both potential changes at the interfaces between the channel and the source/drain, and that with the parameters used, the channel changes from a barrier to a well after the onset of the drain current, because the conduction band edge is below the Fermi level. This is especially true when the spacers act as small barriers in the presence of a gate bias.[14] For simplicity, we assume that there are localized states at $\frac{\hbar^2}{2m_x}\left(\frac{n_x\pi}{l}\right)^2$, neglecting the fact that the potential profile deviates from that of a square quantum well in the presence of an applied gate voltage because of the edge effects. This is qualitatively similar to the simulated results of the ground state and the first two excited states along the channel length direction, obtained with the Schrodinger-Poisson method for a GAAFET.[15] It should be pointed out that the simulated results shown in this study only depend on a few low-lying states. Therefore, the height of the barriers is not expected to qualitatively change the characteristics of the low-lying states that contribute to the current.

With the approximations mentioned above, we have an analytical expression for the energy values of confined states:

$$E_{n_x,n_y,n_z} = \frac{\hbar^2}{2m_x}\left(\frac{n_x\pi}{l}\right)^2 + \frac{\hbar^2}{2m_y}\left(\frac{n_y\pi}{w}\right)^2 + \left[\frac{3}{2}\pi\left(n_z - \frac{1}{4}\right)\right]^{\frac{2}{3}}\left[\frac{(eF\hbar)^2}{2m_z}\right]^{\frac{1}{3}} \qquad (2)$$

where $n_x, n_y$ and $n_z = 1, 2, 3, ..., l$ is the channel length in the x-direction, and $w$ the channel width in the y-direction and the gate field along the z-direction. The 2-fold valleys have the same $m_x$ and $m_y$ values along the channel length and width. The 4-fold valleys also have the same mass along the channel length and width directions, but different from that of the 2-fold valleys.

The confinement energies from Eq. (2) and the bending of the conduction band edge, referred to as the surface potential $\phi_s$, will determine their energy values with respect to the Fermi energy. We derived an expression for the electric field/slop in the triangular well based on Gauss's law and depletion approximation:[9, 16]

$$F = -\frac{1}{2}\frac{-C_{ox}(V_g - V_{fb} - \phi_s) + \sqrt{2\varepsilon_{si} e N_A \phi_s}}{\varepsilon_{si}}, \quad (3)$$

where $C_{ox} = \frac{\varepsilon_{ox}}{t_{ox}}$ the capacitance per unit area associated with the oxide layer, $t_{ox}$ the thickness of the oxide layer, and $\varepsilon_{ox}$ the permittivity of the oxide, $V_g$ the applied gate voltage, $V_{fb}$ the flat-band voltage and $\varepsilon_{si}$ the permittivity of silicon.

The inversion charge $Q_{in}$ is related to the surface potential $\phi_s$,

$$\phi_s = V_g - V_{fb} - \frac{\sqrt{2\varepsilon_{si} e N_A \phi_s}}{C_{ox}} + \frac{Q_{in}}{C_{ox}}. \quad (4)$$

This can be solved analytically with a 2D density of states.[16] Because we are not dealing with a 2D system but a channel with small sizes, we solved the problem using a self-consistent approach, as mentioned earlier.

With the slope of the triangular well and the conduction band edge determined, the energy states can be calculated and are shown in Fig. 2,

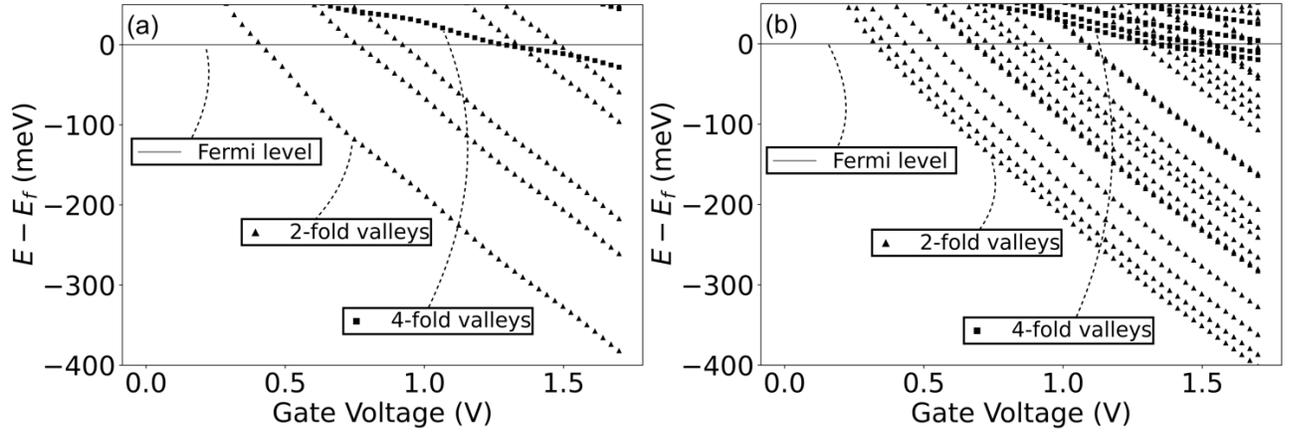

Fig. 2. The energy states with respect to the Fermi level for (a) a channel that is 7 nm long and 6 nm wide at T = 300 K. (b) The energy states for a channel with 7 nm in length and 20 nm in width at T = 300 K.

It is desirable to have large energy separations between the quantized states, relative to the thermal energy, which is the case with the dimensions used in Fig. 2. The anisotropy of the effective mass for different valleys plays an important role. As the triangular well narrows with increasing gate bias, states associated with the first excited state of the triangular well do not drop significantly below the Fermi level. On the other hand, confined states in the square wells, i.e., channel width and length directions associated with the ground state of the triangular well, can move significantly lower than the Fermi level. This is because the shapes of the wells along these directions are less affected by the gate bias. It should also be noted that the calculated quantized states significantly above the Fermi level are less relevant because of the reduced quantization at higher energies. However, such states do not play a significant role in the drain current.

There are several models that have been used for transport within the channel, based on the nature of transport in the channel, i.e., whether it is diffusive or ballistic, or under certain conditions, tunneling. To calculate the drain current when a drain-source voltage is applied, we use the Landauer's formula for ballistic transport because of the small channel length.[17] Hence:

$$I = \frac{e}{L}\sum_n v(E_n)\left[f(E_n - E_{FS}) - f(E_n - E_{FD})\right], \tag{5}$$

where $L$ is the channel length, $E_{FS}$ and $E_{FD}$ the quasi Fermi energies of the source and drain regions, respectively, $E_n$ the energy of the quantized states, $v(E_n)$ the corresponding velocity, and $f(E_n - E_F)$ the Fermi distribution.

In the nonequilibrium situation in the channel with a current, the two quasi-Fermi energies at the source and the drain differ by the amount: $E_{FS} - E_{FD} = eV_{DS}$, where $V_{DS}$ is the applied drain-source voltage. The Fermi energy at the source is taken to be the same as the Fermi energy inside the silicon substrate. We approximated the velocity classically: $v(E_n) = \sqrt{\frac{2E_{n_x}}{m^*}}$, where $m^*$ is the effective mass of the electrons. The calculated velocities of the electrons are on the order of $10^5$ m/s, which is consistent with the reported values.[17]

For this simulation, the applied drain-to-source voltage is 100 $mV$ and the results are shown in Fig. 3. It should be noted that this is done at the low end of the range of drain-source voltage for simplicity, which allows for a first-order perturbation treatment of the quantized states in Fig. 2, rather than completely recalculating these states. A more detailed study is needed to include a more accurate channel potential profile under the gate bias, including the edge effect, and the drain-to-source voltage.

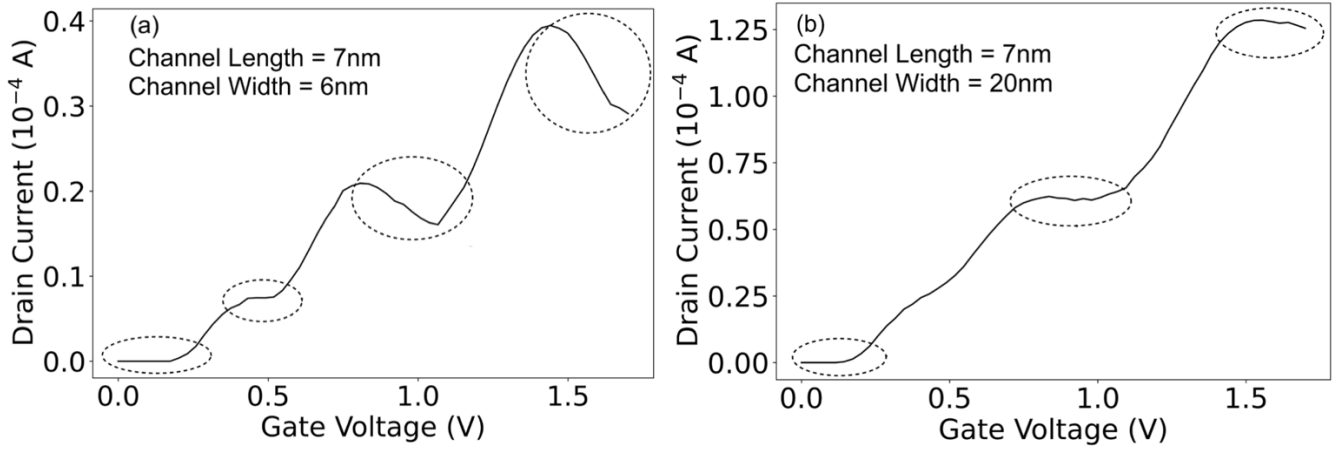

Fig. 3. Drain currents vs. the gate voltage at 300 K for (a) a 7 nm long and 6 nm wide channel and (b) a 7 nm long and 20 nm wide channel.

The currents in Fig. 3 show step-like regions, circled in the plots. While the plots do not give perfect steps, they provide easily identifiable regions at room temperature that can be exploited for multinary logic. It can also be seen that as long as one of the dimensions is 7 nm, step-like features are clear, even for large width. This can be the case in some state-of-the-art devices, including FinFETs that typically have large effective widths, corresponding to the height of the fins. The shapes of the step-like features are determined by the temperature, the source-drain voltage, and the size/shape of the channel and correspondingly if it resembles a 2D, a 1D or a 0D system.

The advantage of the approach discussed here is that it can be realized with the currently used Si technology, and the effect is visible at room temperature. Further studies are needed to more precisely describe the structure used in this report, as already mentioned, and also other structures, such as FinFETs and GAAFETs. While the quantized states in more sophisticated structures have significantly different forms, it can be expected that the localized states in structures with similar dimensions can have energies in the same range, as can be seen from the energy expression, $<\Psi_n|H|\Psi_n>$, where $\Psi_n$ is the wavefunction of a state and highly dependent on the dimensions, and $H$ is the Hamiltonian. As already mentioned, the

short channel effect is not included in this study of size-related quantization. Further studies are needed to investigate FinFETs and GAAFETs, where the short channel effect is greatly reduced. Such studies will be valuable to determining the optimal parameters for specific device designs for applications.

Irrespective of the details of the structures, quantization of electrons will play an important role with decreasing device size, which can be exploited for both processors and floating gate memory devices. As mentioned earlier, the step-like regions in Fig. 3 can be exploited for multinary logic. On the other hand, for the present binary transistors, using the low gate voltage range as 0 and high as 1, the step-like region in the middle of Fig. 3b has to be excluded when assigning the voltage range for the 0 and 1 states. This reduces the voltage variations that can be tolerated.

In the nanometer range, edge effects can be significant for the applied fields from the gate. The shapes of the quantum wells formed by the gate bias with edge effects typically show significant deviations from square wells. The bottom of a well can have the shape close to a parabola. But the quantized energy levels remain in the same energy range. There is another effect in such a case, which is that the well widths are typically reduced, which is a benefit to the demand on small dimensions. Although the detailed dimensions are often proprietary, reductions of the feature sizes are expected to continue in the future, making the effect of quantization within the channel more important. The results show that the device sizes are near the regime where the present device architectures and dimensions can be explored for utilizing quantized states for multinary computation. Moreover, it is advantageous that this approach does not have compatibility problems with the existing technology, which has been one of the problems for several other devices with great promise, such as single electron transistors.